\documentclass[twocolumn,preprintnumbers,amsmath,amssymb]{revtex4}

\usepackage{dcolumn}
\usepackage{bm}

\pdfoutput=1

   \ifx\pdfoutput\undefined
   \usepackage[dvips]{graphicx}
   \else
   \usepackage[pdftex]{graphicx}
   \fi
   \usepackage{epstopdf}
   \DeclareGraphicsExtensions{.pdf,.gif,.jpg,.eps,.png}

\begin{document}

\preprint{{\it Submitted to The Physics of Plasmas}}

\title{The Weibel Instability inside the Electron-Positron Harris Sheet}
 
\author{Yi-Hsin~Liu}
\affiliation{University of Maryland, College Park, MD 20742}
\author{M.~Swisdak}
\affiliation{University of Maryland, College Park, MD 20742}
\author{J.~F.~Drake}
\affiliation{University of Maryland, College Park, MD 20742}

\date{\today}

\begin{abstract}
Recent full-particle simulations of electron-positron reconnection have
revealed that the Weibel instability plays an active role in
controlling the dynamics of the current layer and maintaining fast reconnection.  A four-beam
model is developed to explore the development of the instability within a narrow
current layer characteristic of reconnection.  The problem is reduced
to two coupled second-order differential equations, whose growing
eigenmodes are obtained via both asymptotic approximations and finite
difference methods.  Full particle simulations confirm the linear
theory and help probe the nonlinear development of the instability.
The current layer broadening in the
reconnection outflow jet is linked to the scattering of high-velocity streaming
particles in the Weibel-generated, out-of-plane magnetic field.
 
\end{abstract}

\maketitle

\section{Introduction}
The temperature anisotropy driven Weibel instability \cite{weibel59a}
is thought to play an important role in several astrophysical
systems. For instance, Weibel-mediated collisionless shocks in relativistic jets, pulsar winds, and gamma-ray bursts
have been suggested (\cite{medvedev99a} \cite{gruzinov99a}
\cite{chang08a}) as a possible particle acceleration
mechanism. The Weibel-generated magnetic field scatters particles, enabling them to
bounce back and forth across the shock front, leading to acceleration
via the first-order Fermi mechanism.\cite{spitkovsky08a}

Recently the role of the Weibel instability in electron-positron
(pair) reconnection has begun to receive notice.  Magnetic
reconnection is a fundamental problem in plasma physics, and is
ubiquitous in astrophysical phenomena involving magnetic fields, where
it is the principal mechanism for transforming magnetic energy into
kinetic energy and heat.  Historically, the greatest difficulty in
modeling reconnection has been to demonstrate that it is fast enough
to match observations of energy release in, for instance, solar
flares. By comparing multiple simulation models (e.g., two-fluid,
hybrid, and full particle-in-cell (PIC)) the GEM challenge \cite{birn01a} demonstrated
that inclusion of the Hall term in the generalized Ohm's law was
sufficient to produce fast reconnection.  However, recent studies of
electron-positron reconnection (in contrast to the usual
electron-proton case) raised serious questions about the necessity of
the Hall term for producing fast reconnection.  In contrast with
electron-proton plasmas, the mass symmetry in pair plasmas eliminates
the Hall term.  Yet, simulations suggest pair reconnection is still
fast.  Bessho \& Bhattacharjee \cite{bessho05a} attribute this fact to
contributions from the off-diagonal components of the pressure
tensor. Daughton \& Karimabadi \cite{daughton07a} later discussed the role of 
island formation along the reconnection layer. However Swisdak et al.\cite{swisdak08a} recently
proposed that the Weibel instability, driven by an temperature
anisotropy arising as inflowing plasma mixes with outflow from the
x-line, localizes the reconnection layer, and leads to fast pair reconnection.

\begin{figure}
 \includegraphics[width=8.5cm]{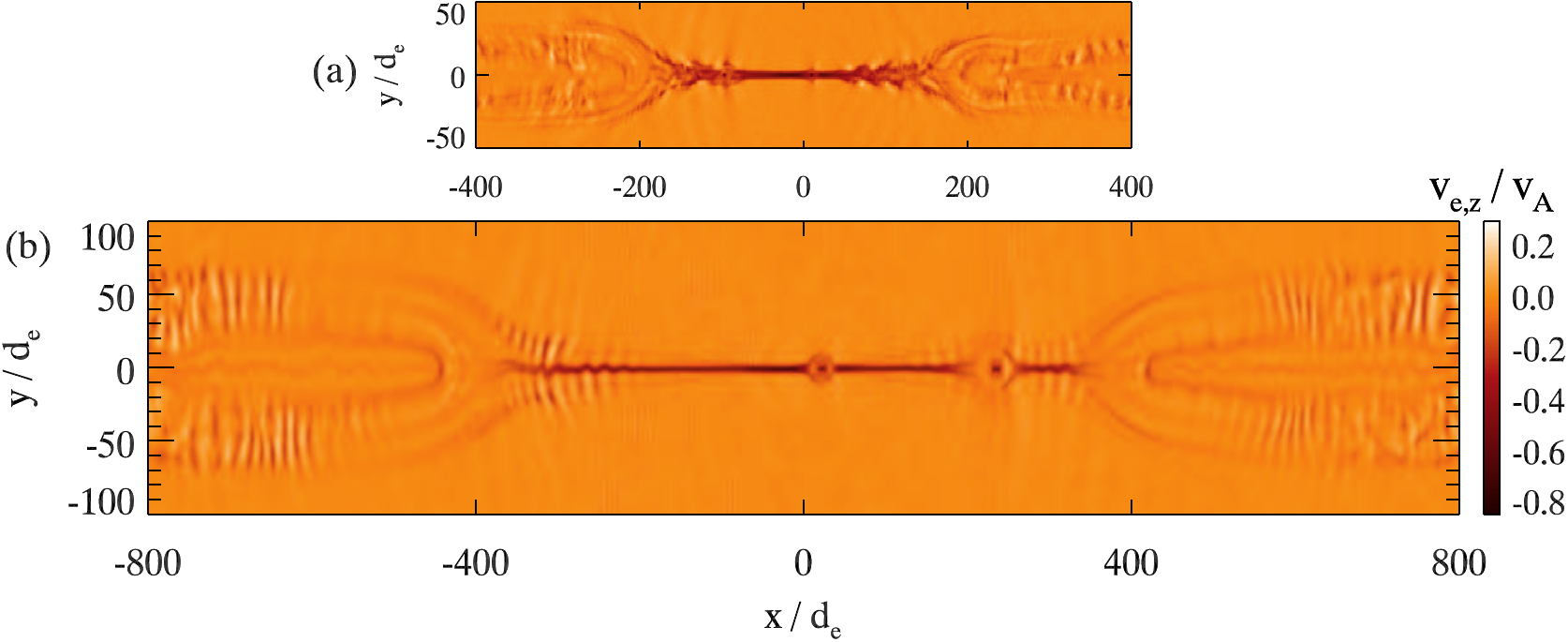} 
 \caption{(Color online) Out-of-plane electron velocity for two pair reconnection simulations. In (a) the current layer length is about $\sim 120 d_e$ (i.e., $|x| \lesssim 60d_e$) \cite{swisdak08a}. In (b) the current layer length of a simulation with $B_z=0$ scales as the system size, $\sim700d_e$ (i.e., $|x| \lesssim 350d_e$). The structure in the large island, which may be due to a two-stream instability, does not affect the behavior near the x-line.}
 \label{1600curr}
\end{figure}

As an example of the importance of the Weibel instability in pair reconnection, we show, in Fig.~\ref{1600curr}, that suppressing the Weibel instability strongly influences the morphology of the current layer. In Fig.~\ref{1600curr}(a) we show a standard pair reconnection simulation where the Weibel instability causes the current layer to become turbulent and broaden, which opens the outflow exhaust as in Petschek's model \cite{petschek64a}. In Fig.~\ref{1600curr}(b), we show the results of a simulation in which we suppress the instability by forcing the out-of-plane component of the magnetic field to zero. 
It is evident that without the turbulence provided by the Weibel instability the narrow current layer extends to the system size. The longer current layer reduces the reconnection rate in the simulation of Fig.~\ref{1600curr}(b) by one third. Since the Weibel instability plays such an important role in maintaining fast reconnection in pair plasma, a thorough understanding of its development in current layers is crucial.

Although Swisdak et al. \cite{swisdak08a} proposed that the Weibel
instability strongly influences reconnection, they only briefly
considered the effects of the current layer environment on the
instability's development. In pair reconnection the outflow layer is
typically narrow (on the order of a few electron inertial lengths) and
confined on both sides by regions of strong magnetic field.  In this
work we take a closer look at the role of the current layer and, in
particular, how it slows (or prevents) the onset of the Weibel
instability. We also try to understand how the unstable Weibel mode is able to open (broaden) the current layer.


In section \ref{seceqns} of this paper we introduce our analytic model
and its assumptions. Section \ref{sechomo} includes the derivation of
the homogeneous dispersion relation and a comparison with kinetic
theory. In Section \ref{secinhomo} we introduce the inhomogeneity
arising from the reconnection geometry, and then numerically compute
the eigenmodes and benchmark them with asymptotic theory in the limits
of large and small current layer widths.  In section \ref{secpic}, we
report on particle simulations that produce Weibel modes inside a
Harris sheet. In section \ref{secimplications} the implications for
pair plasma reconnection are discussed and the downstream turbulent
structure is compared with that of unstable Weibel modes.
  
  \section{The governing equations}\label{seceqns}
 
\begin{figure}
 \includegraphics[width=8.5cm]{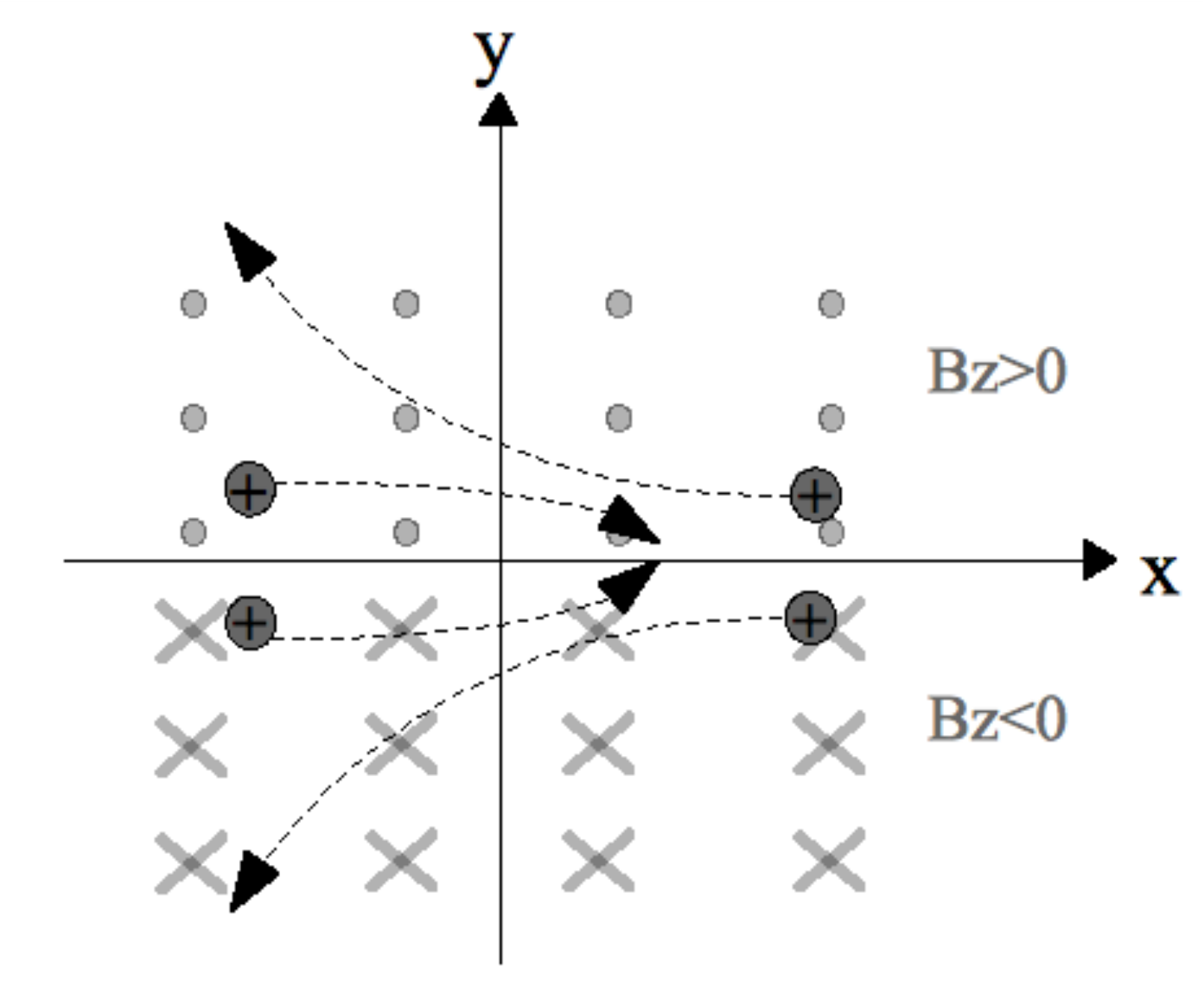} 
 \caption{Schematic of the Weibel instability.}
 \label{weibel_0}
\end{figure}

A cartoon of the basic Weibel instability is shown in
Fig.~\ref{weibel_0}. Consider a uniform unmagnetized plasma with beams
counter-propagating in the x-direction (more generally, possessing an
anisotropic temperature with $T_x> T_y, T_z$).  If a sinusoidal
magnetic field component $B_z(y)$ arises from noise the positively
charged particles with velocities $v_x>0$ will converge towards the
$x$-axis because of the $\mathbf{V \times B}$ Lorentz force while those
with $v_x <0$ will diverge. The combination leads to a current density
$J_x(y)$ of the correct sign to amplify $B_z$, thus driving the mode
unstable.  (Negatively charged particles move in the opposite
directions but have the same net effect.)

To explore the structure of the Weibel instability in reconnection generated current layers, we use a fluid description. Since we are considering a pair plasma we include four species, $\alpha \in \{p+, p-, e+, e-$\}, in our
model: species of positrons and electrons with bulk
velocities $\bf{V}$ and $-\bf{V}$.  In standard notation the governing
equations are: \begin{equation}
\frac{\partial}{\partial t} n_\alpha+\nabla\cdot(n_\alpha {\bf V_\alpha})=0;
\label{eq_full}
\end{equation}
\begin{equation}
  m\frac{d}{dt}{\bf V_\alpha}= q_\alpha {\bf E}+\frac{q_\alpha}{c}{\bf
  V_\alpha} \times\ {\bf B}-\frac{\nabla \cdot \mathrm{P}_\alpha}{n_\alpha};
\end{equation}
\begin{equation} 
 \nabla\times{\bf B}=\frac{4 \pi}{c}\sum_\alpha n_\alpha q_\alpha {\bf
 V_\alpha};
\end{equation}
\begin{equation}  \label{eq_full_end}
  \frac{\partial}{\partial t} {\bf B} = -c  \nabla \times \bf{E}.
\end{equation}
We assume the pressure tensor can be written in the diagonal form
\begin{equation}
\mathrm{P}_{\alpha}=n_{\alpha} T_{\alpha}=n_{\alpha} \left [
     \begin{array}{clr}
     T_{xx, \alpha} & 0 & 0 \\
     0 & T_{yy, \alpha} & 0 \\
     0 & 0 & T_{zz, \alpha}
     \end{array} \right]
\end{equation} 
and that the temperature components do not vary in space or time.  

To particularize our coordinates we take the counter streaming velocities to be parallel to $\mathbf{\hat{x}}$.  The
perturbed magnetic field of the Weibel mode can then, without loss of
generality, be taken to be parallel to $\mathbf{\hat{z}}$ and the
wavevector parallel to $\mathbf{\hat{y}}$.  All physical quantities
are assumed uniform in both the $x$ and $z$ directions,
$\partial/\partial_z=\partial/\partial_x=0$.

Our initial state is characterized by 
\begin{equation}
\begin{split}
{\bf{V}}= V_x(y)\mathbf{\hat{x}}; \qquad {\bf{B}} = B_x(y)\mathbf{\hat{x}};\\
 n_\alpha=n(y); \qquad T_{yy,\alpha}=T_{zz,\alpha}=T.
\end{split}
\end{equation} 
Pressure balance requires $\left(4nT_{yy}+B_x^2/8\pi \right) '=0$, where a prime stands for $\partial/\partial y$, and the total number density for either electrons or positrons is $2n$.
If all perturbed variables are proportional to $e^{\gamma t}$ we
can linearize equations (\ref{eq_full})-(\ref{eq_full_end}) as
\begin{equation} \gamma \tilde{n}+ n' \tilde{V}_{y,\alpha}+n
\tilde{V}_{y,\alpha}'=0; \\ \label{linear1} \end{equation}
    \begin{equation}
  \gamma m \tilde{V}_{x,\alpha}+m \tilde{V}_{y,\alpha} V'_{x,\alpha}=q_{\alpha} \tilde{E}_x;
   \label{linear2}
  \end{equation}
   \begin{equation}
  \gamma m \tilde{V}_{y,\alpha}=q_{\alpha}
  \tilde{E}_y+\frac{q_{\alpha}}{c}[\tilde{V}_{z,\alpha} B_x-V_{x,\alpha}
  \tilde{B}_z]+T\frac{n'}{n^2}\tilde{n}-\frac{T}{n}\tilde{n}';
  \label{linear3}
  \end{equation}
  \begin{equation}
  \gamma m \tilde{V}_{z,\alpha}=q_{\alpha} \tilde{E}_z-\frac{q_{\alpha}}{c} \tilde{V}_{y,\alpha} B_x;
  \label{linear4}
  \end{equation}
  \begin{equation}
  \tilde{B}_z'=\frac{4 \pi}{c}
  \sum_{\alpha} q_{\alpha} (\tilde{n}V_{x,\alpha}+n \tilde{V}_{x,\alpha});
  \label{linear5}
  \end{equation}
  \begin{equation}
  \gamma \tilde{B}_z=c \tilde{E}_x',
  \label{linear6}
  \end{equation}
where a tilde indicates a perturbed quantity.

The effective x-direction temperature is $T_{x} =mV_{x}^2+T_{xx}$.
Note that by assuming $\partial/\partial x = 0$ we eliminate $T_{xx}$
from the equations. The effective temperature anisotropy is then
determined only by the streaming velocity.

Due to the mass and charge symmetry between electrons and positrons, we
can collapse these eighteen equations,
Eq.~(\ref{linear1})-Eq.~(\ref{linear6}), into two second-order
differential equations (see Appendix A for details):
\begin{equation}
C_s^2{\bar{\chi}}''-C_s^2\frac{n'}{n} \bar{\chi}'-\Omega^2\bar{\chi}-
\frac{4e}{m}V_{x}n\tilde{E}_x' =\gamma^2\bar{\chi};
 \label{eq1}
\end{equation}
 \begin{equation}
 \tilde{E}_x''-\frac{2}{d^2}\tilde{E_x}+\frac{4\pi
 e}{c^2}(V_{x}\bar{\chi})'=0
 \label{eq2}
\end{equation}
with the following definitions: $\bar{\chi} \equiv n
(\tilde{V}_{y,p+}-\tilde{V}_{y,p-}-\tilde{V}_{y,e+}+\tilde{V}_{y,e-})$; $\Omega$ is the
gyrofrequency based on $B_x$; $ C_s \equiv \sqrt{T/m}$ is the sound speed; and $d \equiv \sqrt{mc^2/8 \pi n e^2}
\equiv c/\omega_p$ is the skin depth with $\omega_p$ the plasma
frequency.  

\section{The Weibel instability in a homogeneous plasma}\label{sechomo}

In a uniform magnetized plasma a dispersion relation can be found by
combining Eqs.~(\ref{eq1}) and (\ref{eq2}) and letting
$\partial/\partial_y \rightarrow ik_y$:
\begin{equation}
  \gamma^2=\frac{2V_{x}^2k_y^2}{2+k_y^2d^2}-C_s^2k_y^2-\Omega^2.
  \label{homo_disp}
\end{equation}
Clearly while the streaming temperature (the first term of the right
hand side) serves as the instability driver, thermal effects in the
y-direction (the second term) and the background magnetic field (the
third term) stabilize the mode.

In the even simpler case of a strongly anisotropic unmagnetized
plasma, the kinetic growth rate of the Weibel instability is
\cite{swisdak08a},\cite{krall86a}:
\begin{equation}
\gamma^2 \approx  \frac{2 v^2_{th,x} k_y^2}{2+k_y^2d^2},
\label{Krall}
\end{equation}
where $v_{th,x}=\sqrt{T_x/m}$ is the thermal velocity in the
$x$ direction.  It is evident that this matches the first term of
Eq.~(\ref{homo_disp}) with $v_{th,x}\rightarrow V_{x}$. The validity of
using counter streaming cold plasma beams to analyze a single warm
plasma was demonstrated by Davidson et al. \cite{davidson72b}, who
showed that this instability is not affected by the detailed shape of
the plasma distribution function, but only by the effective
temperature.

\section{The Weibel instability inside a Harris sheet}\label{secinhomo}

The results of Section \ref{sechomo} were derived in the context of a
homogeneous plasma.  However for reconnection simulations it is
important to study how the development of the Weibel instability
proceeds inside a narrow current layer.  In this section we seek to
understand whether, and to what degree, the inhomogeneities in the
plasma density and background magnetic field affect the instability.

\begin{figure}
 \includegraphics[width=8.5cm]{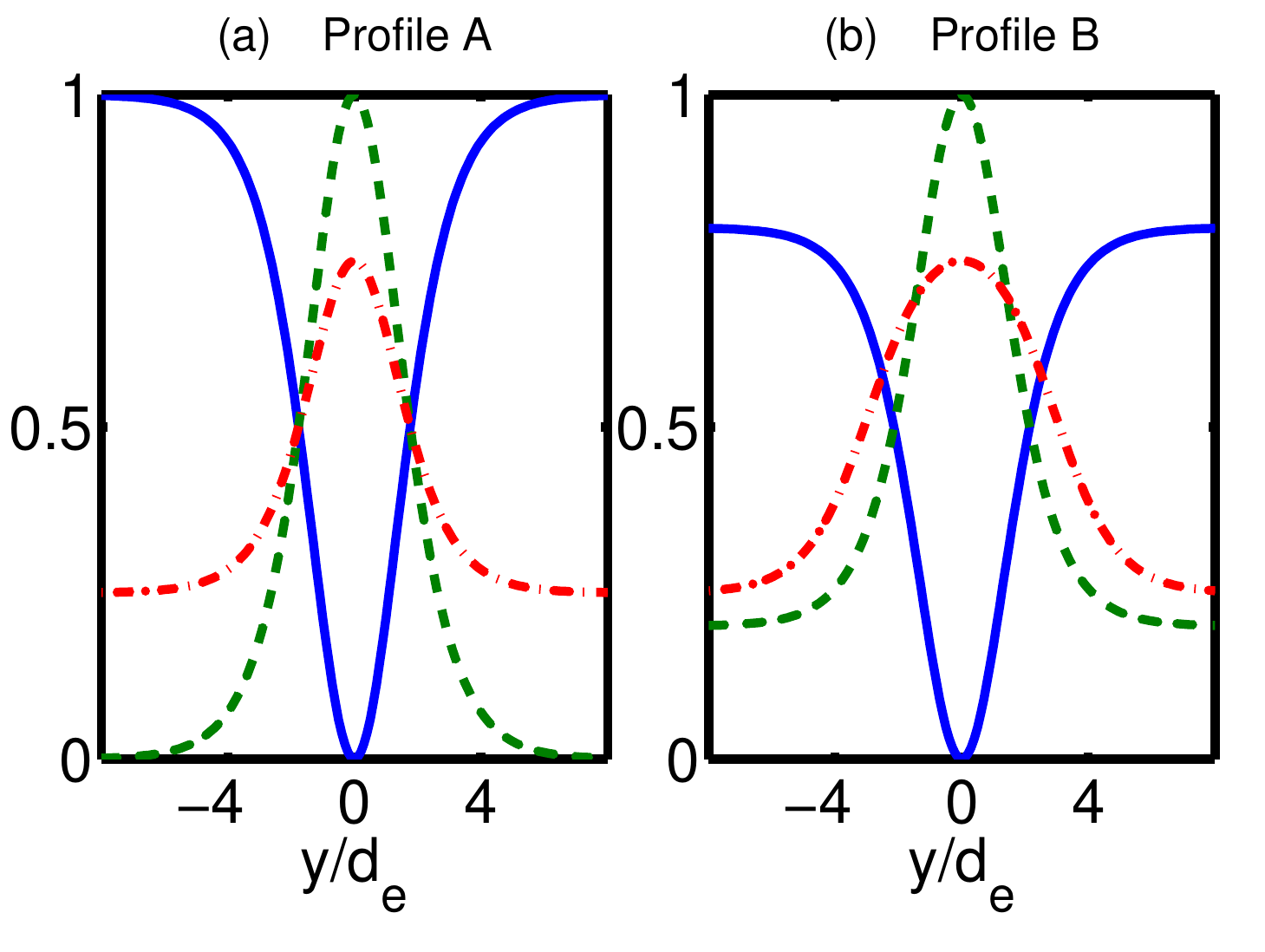} 
 \caption{(Color online) Current sheet profiles.  (Blue) solid curves
 are $B_x^2$, (green) dashed curves are the positron/electron density $n_{p,e}$, 
 and (red) dot-dashed curves are x-direction temperature $T_x$.  Profile A in (a) with
 $n_0=1.0n_0$, $B_{x0}=1.0B_{x0}$, and $\delta =2d_e$. Profile B in (b) with $n_h=0.8n_0, n_b=0.2n_0,
 B_{x,h}^2=0.8B_{x0}^2$ and $\delta=2d_e$.} 
 \label{inhomo}
 \end{figure}

Our equilibrium is taken to be the usual Harris layer with a slight
modification that incorporates an anisotropic plasma temperature
\begin{equation}
 B_x=B_{x0}\tanh(y/\delta); \qquad n_{p,e}=2n =n_0\mbox{sech}^2(y/\delta);
 \label{inhomo_th1}
\end{equation}
\begin{equation}
 V_x^2=(V_{x0}^2-C_s^2)\mbox{sech}^2(y/\delta)+C_s^2,
\label{inhomo_th2}
\end{equation}
where $B_{x0}, n_0$, and $V_{x0}$ are constants, $\delta$ measures the
width of current sheet and the subscripts p/e stand for positron/electron.  We will refer to this setup as
Profile A. The profiles are shown in Fig.~\ref{inhomo}(a).

For the sake of comparing with full particle pair reconnection
simulations, all physics quantities are presented in the same
normalized units as those in Swisdak et al. \cite{swisdak08a}: the
magnetic field to the asymptotic value of the reversed field $B_{x0}$, the
density to $n_0$ which is the value at the center of the current sheet minus a
possible uniform background density, velocities to the electron
Alfv\'en speed $V_{A,e}$, lengths to the electron inertial length
$d_e\equiv \sqrt{mc^2/4\pi n_0 e^2}$, times to the inverse electron cyclotron frequency
$\Omega_{ce}^{-1}$, and temperatures to $m_e V_{A,e}^2$. We use $T=0.25 m_e V_{A,e}$ in the analysis of this
section, which is the initial set up of the pair reconnection
simulation of \cite{swisdak08a}, except for the latter's inclusion of a uniform
background density of $0.2 n_0$.

To find the modes of Eqs.~(\ref{eq1}) and (\ref{eq2}) we discretize the
governing equations in the $y$ direction (imposing zero derivative
boundary conditions) and find the eigenvalues of the resulting matrix.
We use a grid size of $\delta/100$ and domain size of $20 \delta$,
both of which are sufficient to ensure covergence.  Before describing
the numerical results, however, we investigate the behavior of the equation analytically. 

\subsection{$k_yd_e\gg 1$, $\delta/d_e \ll 1$}

In this limit the current layer thickness and the wavelength of the
instability are much smaller than the inertial length $d_e$.  From
general considerations we expect the mode to be harder to excite in
such circumstances, unless $V_x$ is large.  It is straightforward to
combine Eqs.~(\ref{eq1}) and (\ref{eq2}) into a single second-order
ordinary differential equation by eliminating the second term in
Eq.~(\ref{eq2}), which is small since $k_yd_e\gg 1$. The result is
\begin{equation}
C_s^2\bar{\chi}''-C_s^2\frac{n'}{n}\bar{\chi}'+\left(\frac{2V_{x}^2}{d^2}
- \Omega^2\right)\bar{\chi}=\gamma^2\bar{\chi}.
\label{bigk}
\end{equation}
For the parameter regime of interest, $d/d_e \sim \Omega/\Omega_{ce} \sim O(1)$.
Therefore for every term in Eq.~(\ref{bigk}) to be of the same
order (except, perhaps, the $\Omega^2$ term), the following scaling rules
must apply:
\begin{equation}
 \gamma \sim V_x/d \sim k_y C_s \gg \Omega \sim C_s/d.
 \label{scale_largek}
\end{equation}
After substituting for the functional form of the  inhomogeneity and
changing variables to $u=\cosh(y/\delta)\bar{\chi}$, we have
\begin{equation}
  C_s^2 u''=Q(y)u,
  \label{WKB}
\end{equation}
where 
\begin{equation}
\begin{split}
Q=\gamma^2
-\frac{2}{d_e^2}\left[V_{x0}^2\mbox{sech}^2(y/\delta)+C_s^2\tanh^2(y/\delta)\right]\mbox{sech}^2(y/\delta)\\
+\Omega_0^2\tanh^2(y/\delta)+\frac{C_s^2}{\delta^2}
\end{split}
\label{QQ}
\end{equation}
and $\Omega_0 \equiv B_{x0}e/mc$. For $V^2_{x0} > C_s^2d_e^2/2\delta^2$,
$Q$ is negative near $y=0$ and positive at large $y$ and Eq.~(\ref{WKB}) therefore 
has bounded solutions.

We further simplify $Q$ by Taylor
expanding for small $y/\delta$, and neglecting the $C_s^2 \tanh^2$ and
$\Omega_0^2\tanh^2$ terms in Eq.~(\ref{QQ}), which are small in the ordering given in Eq.~(\ref{scale_largek}),
\begin{equation}
C_s^2
u''-\frac{2V_{x0}^2}{d_e^2\delta^2}y^2u=\left(\frac{2V_{x0}^2}{d_e^2}-\frac{C_s^2}{\delta^2}-\gamma^2\right)u.
\end{equation}
This equation can be solved in terms of Hermite polynomials with
eigenvalues that give a maximum growth rate of
\begin{equation}
\gamma_{max}^2=\frac{2V_{x0}^2}{d_e^2}-\frac{\sqrt{2}V_{x0}C_s}{d_e \delta}-\frac{C_s^2}{\delta^2}.
\label{hermite_large}
\end{equation}
Clearly the density shear term, $C_s^2/\delta^2$, which arises from the second term of Eq.~(\ref{bigk}), has the strongest stabilizing effect.
By letting $\gamma_{max}=0$, we obtain the critical temperature anisotropy
required for the unstable mode in the small $\delta/d_e$ limit,
\begin{equation}
\left(\frac{T_{x0}}{T}\right)_{c}=\frac{3+\sqrt{5}}{4} \left(\frac{d_e}{\delta} \right)^2,
\label{anisotropy_large}
\end{equation}
where $T_{x0} \equiv (T_{x})_{y=0}=m V_{x0}^2$.

\begin{figure}
    \includegraphics[width=8.5cm]{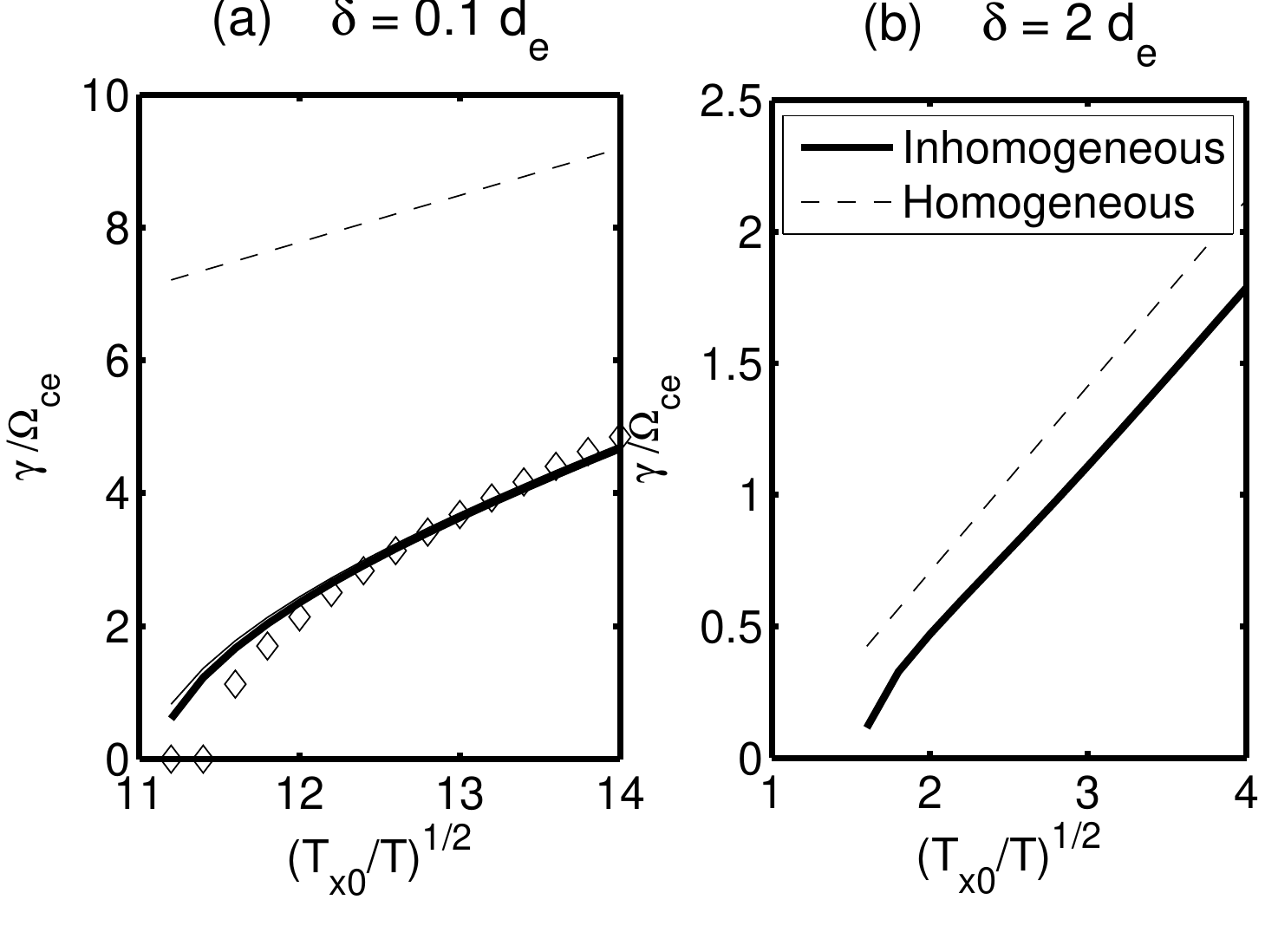} 
\caption{Maximum growth rate
    versus temperature anisotropy for Profile A for (a)
    $\delta=0.1d_e$ and (b) $\delta=2 d_e$.  Thick solid
    curves: numerical solutions of the full four-beam model,
    Eqs.~(\ref{eq1})-(\ref{eq2}). Thin solid curve (only in (a)):
    numerical solutions of the reduced equation, Eq.~(\ref{bigk}).  
    Diamonds (only in (a)): analytical solutions in the large-k limit,
    Eq.~(\ref{hermite_large}).  Dashed curves: analytical solutions for
    an unmagnetized homogenous plasma, Eq.~(\ref{homo_disp}).} 
    \label{gama}
\end{figure}

We plot $\gamma$ versus anisotropy in Fig.~\ref{gama}(a) for
$\delta=0.1 d_e$ and in Fig.~\ref{gama}(b) for $\delta=2 d_e$. In the
$\delta=0.1 d_e$ case, the full equations
(Eqs.~(\ref{eq1}),(\ref{eq2})) and the reduced equation
(Eq.~(\ref{WKB})) result in the same curve because the wavenumber of
the growing mode is large enough to validate the approximation. The
analytical solution from Eq.~(\ref{hermite_large}) follows the correct
trend and matches the numerical results, particularly in the large anisotropy
limit. We observe that $\gamma$ is proportional to $V_{x0}$ (i.e.,
$V_{x0}\sim(T_{x0}/T)^{1/2}$ ) as the scaling rule
(Eq.~(\ref{scale_largek})) suggests. By comparing the growth rate of
the unmagnetized homogeneous plasma from Eq.~(\ref{homo_disp}), it is
also clear that the instability is severely suppressed by the
inhomogeneity.  In contrast, the $\delta=2 d_e$ case results in a
somewhat closer match because of the
increase in the inhomogeneity scale length.

\subsection{$k_yd_e \ll 1$, $\delta/d_e \gg 1$}

Although in this limit the search for bounded modes is rather
complicated, it is possible to gain some insight by expanding the
homogeneous dispersion relation, Eq.~(\ref{homo_disp}) in the small
$k_y$ limit:
\begin{equation}
  \gamma^2 \approx V_x^2 k_y^2 \left(1- \frac{d^2}{2} k_y^2\right) - C_s^2 k_y^2 -\Omega^2.
\end{equation}
If we only keep terms of $O(k_y^2)$ we find that
\begin{equation}
 k_y^2 \approx (\gamma^2+\Omega^2)/(V_x^2-C_s^2),
\end{equation}
which is always positive and has no bounded modes. It is only by
keeping the next term (i.e., $O(k_y^4)$) in the expansion that bounded modes exist, a
fact that will guide our treatment of the full system.

We begin by neglecting the $\tilde{E}_x''$ term in Eq.~(\ref{eq2}),
solving for $E_x$, and then substituting the result back into the
equation, ultimately giving
\begin{equation}
\frac{2}{d^2}\tilde{E_x} \simeq \frac{4\pi
e}{c}(V_{x}\bar{\chi})'+ \frac{4\pi
e}{c}\left(\frac{d^2}{2}(V_x\bar{\chi})' \right)''.
\end{equation}
We then use this approximation in Eq.~(\ref{eq1}) in the small
$y/\delta$ limit and assume that we are close to marginal stability. The resulting equation is: 
\begin{equation}
\frac{V_{x0}^2d_e^2}{2}
 \bar{\chi}''''+(V_{x0}^2-C_s^2) \bar{\chi}''+
 \frac{\Omega_0^2}{\delta^2}y^2\bar{\chi}=-\gamma^2\bar{\chi}.
\end{equation}
The parameter regime of interest to us is $V_{x0}/V_{A,e} \sim C_s/V_{A,e} \sim
\Omega_0/\Omega_{ce} \sim O(1)$.  By requiring each term in this equation to be
the same order we arrive at the ordering
\begin{equation}
\gamma^2 \sim k_y^2 (V^2_{x0}-C^2_s); \qquad k_y^2d_e^2 \sim (V^2_{x0}-C^2_s)/V^2_{x0} \ll 1.
\label{scale_smallk}
\end{equation}
To proceed, we transform the equation to Fourier space
\begin{equation}
\frac{\Omega_0^2}{\delta^2}\frac{\partial^2}{\partial k_y^2}\bar{\chi}+\left[(V_{x0}^2-C_s^2)k_y^2-
\frac{V_{x0}^2 d_e^2}{2} k_y^4\right]\bar{\chi}=\gamma^2 \bar{\chi}.
\end{equation}
The quantity inside the square brackets has a maximum at $k_{y0}=\sqrt{V_{x0}^2-C_s^2}/d_eV_{x0}$.
Since we are looking for the maximally growing mode, we Taylor expand
this quantity in $s\equiv k_y-k_{y0}$ around $s=0$:
\begin{equation}
\frac{\Omega_0^2}{\delta^2}\frac{\partial^2}{\partial s^2}\bar{\chi}-2(V_{x0}^2-C_s^2)s^2\bar{\chi}
= \left[\gamma^2-\frac{(V_{x0}^2-C_s^2)^2}{2d_e^2V_{x0}^2}\right]\bar{\chi}.
\end{equation}
The solution of this equation can again be written in the form of
Hermite polynomials with a maximal eigenvalue of
\begin{equation}
\gamma_{max}^2=\frac{(V_{x0}^2-C_s^2)^2}{2d_e^2V_{x0}^2}-
\frac{\Omega_0}{\delta}\sqrt{2(V_{x0}^2-C_s^2)}.
\label{hermite}
\end{equation}
Without the second term, which arises from the background magnetic
field, the growth rate scales as $\gamma_{max}^2 \sim
(V_{x0}^2-C_s^2)k_y^2$ (with $k_y^2$ given in Eq.~(\ref{scale_smallk})), which is essentially the same result as the
unmagnetized homogeneous relation, Eq.~(\ref{homo_disp}), in the
$k_yd_e\ll 1$ limit.
From Eq.~(\ref{hermite}), we derive the marginal criterion in the large
$\delta/d_e$ limit,
\begin{equation}
\left(\frac{T_{x0}}{T}\right)_c=1+\left(\frac{2\sqrt{2}\Omega_0 d_e^2
}{C_s\delta}\right)^{2/3}.
\label{anisotropy_small}
\end{equation}

\begin{figure}
 \includegraphics[width=8.5cm]{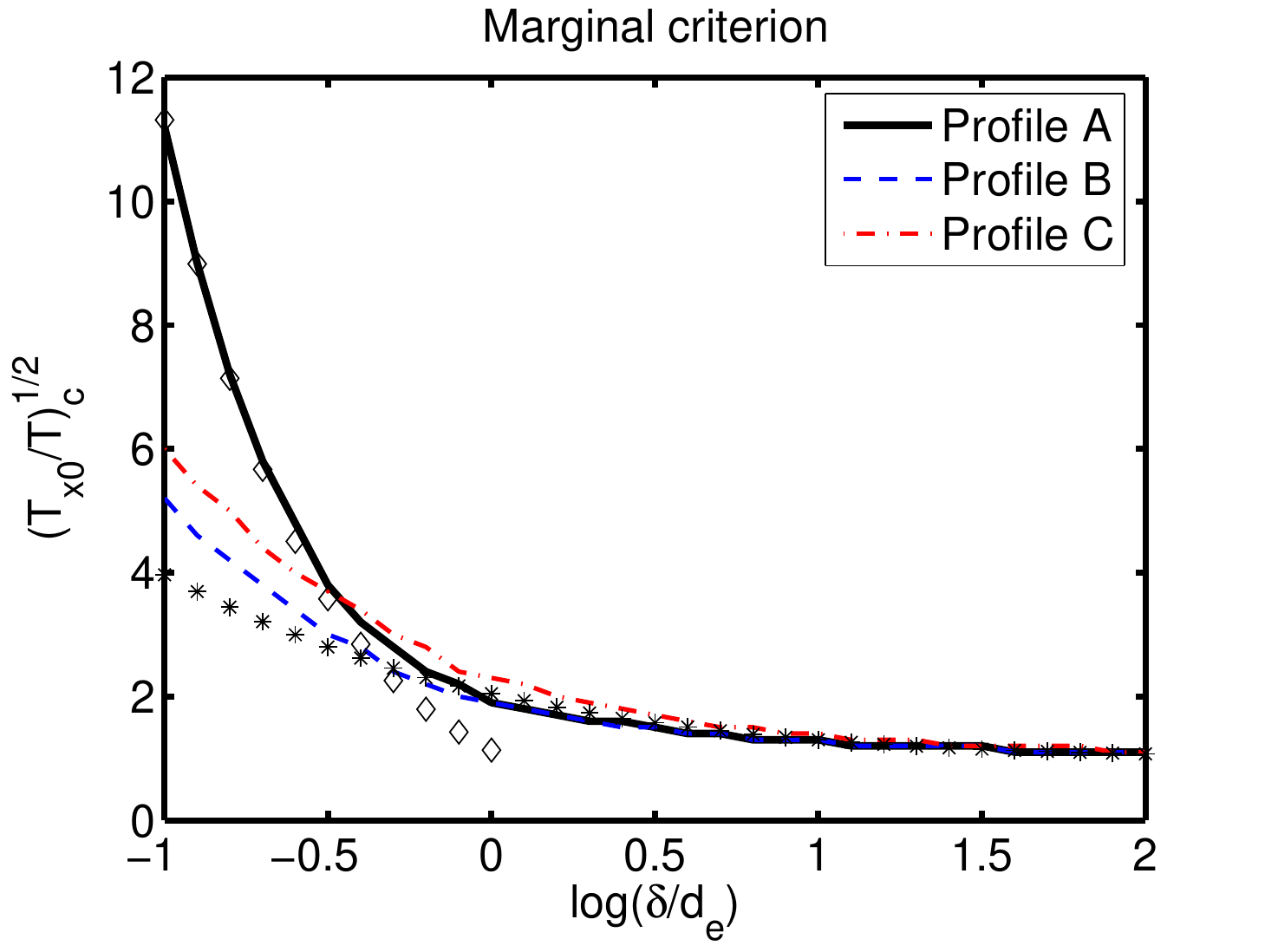} 
\caption{(Color online) The marginal threshold of the Weibel instability 
 as a function of $\delta$. Profile A: the solid curve from the numerical solutions of the full
 four-beam model (Eqs.~(\ref{eq1})-(\ref{eq2})), the diamonds
 from the analytical solution in the small-$\delta$ limit
 (Eq.~(\ref{anisotropy_large})), and stars from the
 analytical solution in the large-$\delta$ limit
 (Eq.~(\ref{anisotropy_small})). Profile B: the (blue) dashed curve from numerical solutions of the
 full four-beam model.  Profile C: the (red) dot-dashed curve from numerical
 solutions of the full four-beam model.} \label{marginal}
\end{figure}

Finally we plot the threshold of marginal instability for different
values of $\delta$ in Fig.~\ref{marginal}. The numerical solution of
our model with Profile A is carefully benchmarked by these asymptotic
theories in both the small and large $\delta$ (or $k_y$) limits. The
(blue) dashed and (red) dot-dashed lines are discussed later.

\section{The small box PIC simulation}\label{secpic}

In order to confirm our linear theory and study the nonlinear
development of the Weibel instability, we conduct several simulations
with the particle-in-cell (PIC) code p3d \cite{zeiler02a}.  The
electromagnetic fields are defined on gridpoints and advanced in time
with an explicit trapezoidal-leapfrog method using second-order
spatial derivatives.  The Lorentz equation of motion for each particle
is evolved by a Boris algorithm where the velocity $\mathbf{v}$ is
accelerated by $\mathbf{E}$ for half a timestep, rotated by
$\mathbf{B}$, and accelerated by $\mathbf{E}$ for the final half
timestep.  To ensure that $\bm{\nabla \cdot} \mathbf{E} = 4\pi \rho $
a correction electric field is calculated by inverting Poisson's
equation with a multigrid algorithm.  Although the code permits other
choices, we work with fully periodic boundary conditions.

We consider a system periodic in the $x-y$ plane. The simulations presented here are two-dimensional, 
i.e., $\partial/\partial z = 0$. The initial
equilibrium consists of a Harris current sheet superimposed on an
ambient population of uniform density $n_b$, 
\begin{equation}
 B_x=B_{x,h}\tanh(y/\delta); \qquad n_{p,e}=n_h\mbox{sech}^2(y/\delta)+n_b, 
 \label{inhomo_pic1}
\end{equation}
where $B_{x, h}, n_h, n_b $ are constants, subscript $h$ stands for Harris sheet and
$\delta$ is the half width of the current sheet. We
input the Harris plasma with an initial temperature $T_{xx, h}
>T_{yy}=T_{zz}\equiv T$.  The background plasma has an isotropic
temperature $T$. Therefore, 
\begin{equation}
T_x=\frac{n_b T+n_h \mbox{sech}^2(y/\delta) T_{xx, h}}{n_b+n_h
\mbox{sech}^2(y/\delta)}, 
\label{inhomo_pic2}
\end{equation}
where $T$ and $T_{xx, h}$ are constants. This equilibrium is very similar to Profile A defined in
Eqs.~(\ref{inhomo_th1})-(\ref{inhomo_th2}) except for the inclusion of
a background density.  In the
following we take the parameters $n_h=0.8n_0, n_b=0.2n_0, B_{x,h}^2=0.8 B_{x0}^2, \delta=2d_e$ in order to compare later to our full simulations of pair reconnection, and
refer to the initial condition as Profile B (see Fig.~\ref{inhomo}(b)).

In order to prevent the simultaneous growth of the tearing mode
\cite{furth63a} we consider a domain with $L_x=4d_e\ll L_y=50d_e$.  
We let the Weibel mode grow from noise for different
values of the temperature anisotropy, measure the growth rate, and
compare the results with the theoretical values from the full
four-beam model in Fig.~\ref{smallrun}(b).  Kinetic effects only slightly reduce the
growth rates of our four-beam results and the eigenfunction predicted by the model is in good
agreement with the simulations (see Fig.~\ref{smallrun}(a)). The
anisotropy threshold for the four-beam model with Profile B is shown as a (blue) dashed line in
Fig.~\ref{marginal}.

The evolution of the temperature anisotropy and Weibel-generated $B_z$
from four small-box runs with different initial conditions is shown
in Fig.~\ref{smallrun}(c)-(d). The anisotropies (at $y=0$) decrease in
time to slightly above the marginal value $\sim 2.3$, while the amplitude of $B_z$
simultaneously rises. The increase of $B_z$
scatters the hot streaming plasma, reducing the central anisotropy. Only a small part of the energy
released transfers to $B_z$. The scattering increases $T_{yy}$, and therefore the central pressure, causing the layer to expand. As a result, the ambient $B_x$ increases due to compression (not shown).  

In homogeneous plasmas we can predict the saturation level by analyzing the particle motion in the y-direction,
$dV_y/dt =-(e/mc)V_x B_z$.
Roughly speaking, the Weibel instability saturates when the magnetic field grows to a value such that the particles become magnetically trapped and can no longer amplify the field. Trapping occurs when the particle excursion along $\hat{y}$, $\Delta y$, is comparable to the wavelength during the mode growth time,
\begin{equation}
  \Delta y \sim \frac{eV_x B_z}{mc\gamma_{max}^2} \sim \frac{1}{k_{max}}.
\end{equation}
Equivalently, 
\begin{equation}
  \gamma_{max} \sim \left(\frac{e}{mc}V_x B_{z} k_{y, max}\right)^{1/2} =\omega_B,
\end{equation}
The saturation occurs when the magnetic bounce frequency, $\omega_B$, is comparable to the fastest linear growth rate. This is essentially the empirical result from Davidson et al.\cite{davidson72b}. In the strong anisotropy limit, $\gamma_{max} \sim V_x k_{y,max}$ (Eq.~(\ref{homo_disp})), we obtain the simple saturation criterion $k_{y,max} \rho_z \sim 1$, where $\rho_z$ is the gyro-radius in the $B_z$ field.
In our inhomogeneous plasmas the same principle
guides the saturation of the Weibel mode, although the predicted
saturation value of $B_z$ is smaller due to the inhomogeneity.

\section{Implications for Pair Reconnection}\label{secimplications}

\subsection{$k_y$ structure}  

\begin{figure}
    \includegraphics[width=8.5cm]{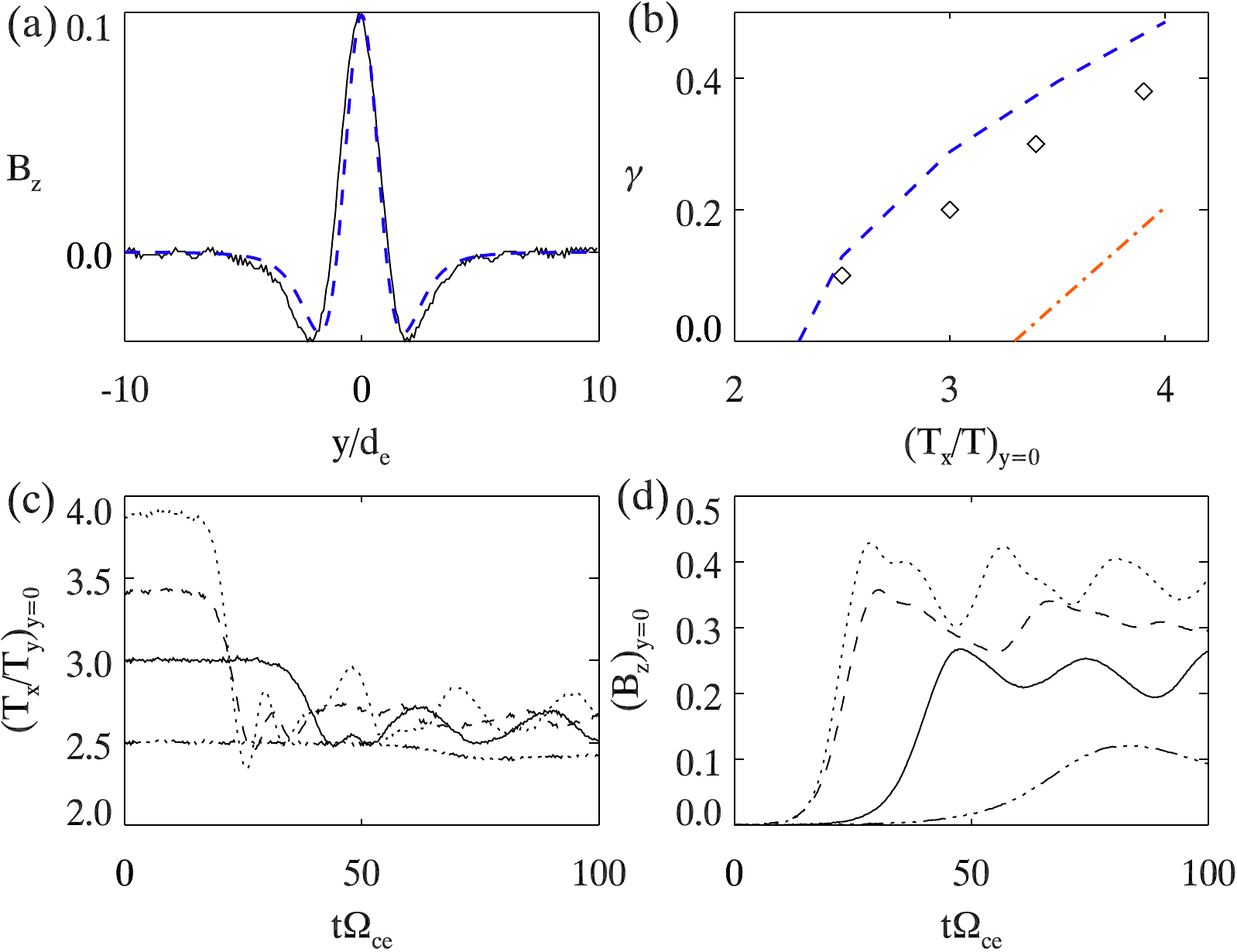} 
\caption{(Color online) Results from 
     PIC simulations with initial Profile B. In (a) the
    eigenfunction of $B_z$ with $T_{x0}/T=3.0$: in
    solid the small-box PIC simulation and in (blue) dash the full four-beam model.  In (b) the maximum growth
    rate versus anisotropy: in diamond the small-box PIC simulations, in (blue) dash the full four-beam model and in (red) dot-dash the full four-beam model with Profile C.
    In (c) the temporal evolution of different initial temperature anisotropies in the small-box PIC
    simulation.  In (d) the magnitude of $B_z$ (at $y=0$) for the
    temperature anisotropies plotted in (c).} \label{smallrun}
\end{figure}

\begin{figure}
    \includegraphics[width=8.5cm]{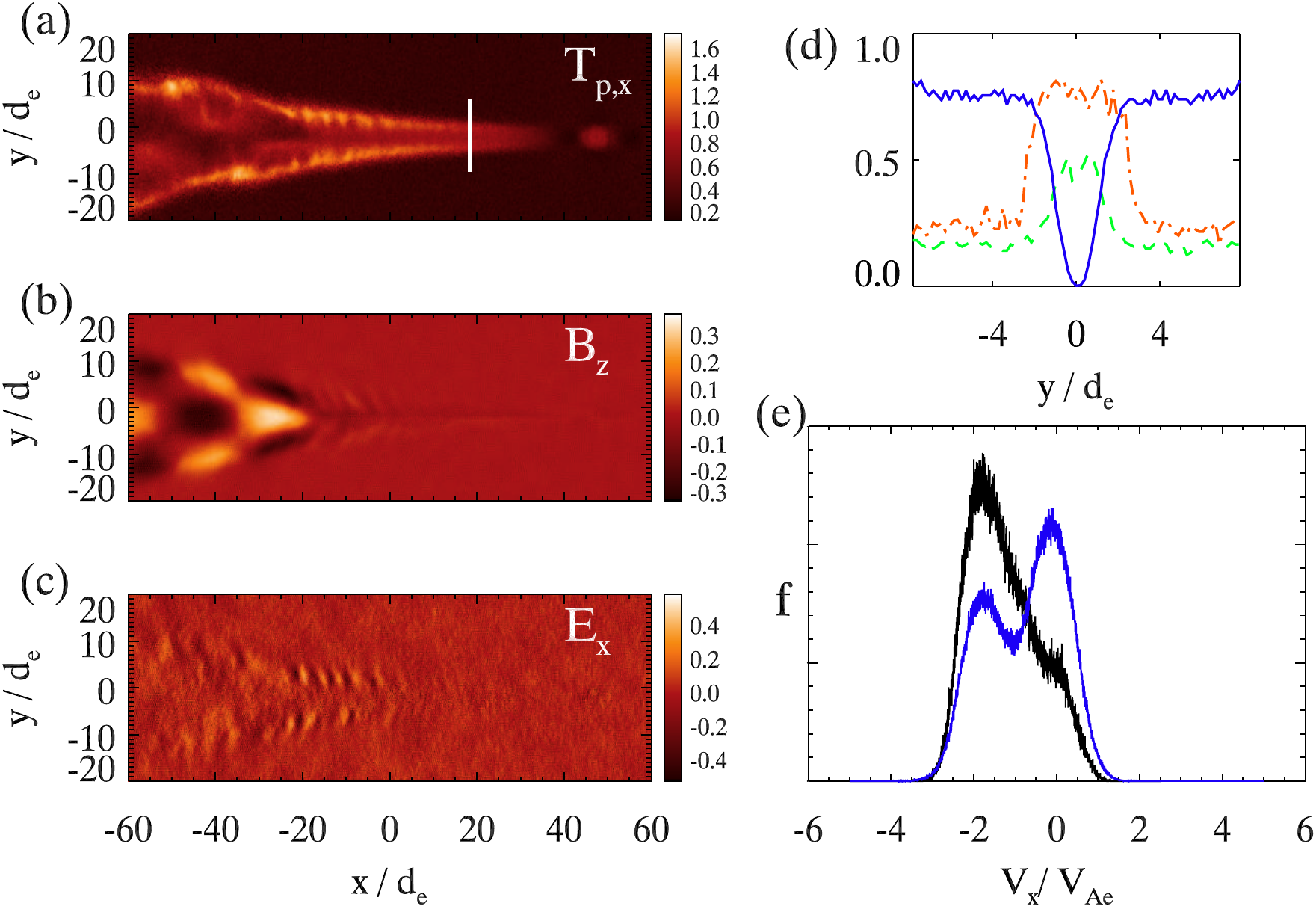} 
\caption{(Color online) A PIC simulation of pair reconnection. In (a) the
    x-direction positron temperature. The X-point is on
    the right edge of the plot.  In (b) the $B_z$ signatures of both the
    Weibel (chess-board-like structure) and two-stream (finer
    structure upstream) instabilities. In (c) the $E_x$ signature of the two-stream
    instability in the downstream region. In (d) the inhomogeneity plotted
    along the white line in (a).  The line styles (colors) are the same as in
    Fig.~\ref{inhomo}.  In (e) the double-humped velocity
    distribution function (blue) in the two-stream and Weibel unstable
    region ($x/d_e,y/d_e \in (-21:18, -4:4)$) becomes single-humped
    (black) farther downstream ($x/d_e,y/d_e \in (-44:-30, -4:4)$). } \label{weibel_2stream}
\end{figure}

The basic features of a pair plasma reconnection simulation are shown in Fig.~\ref{weibel_2stream}. 
The Weibel instability manifests itself as a chess-board-like structure in the downstream out-of-plane magnetic
field (see Fig.~\ref{weibel_2stream} (b)). The structure travels
downstream with the outflow speed from the x-line, which implies that
the Weibel is a purely growing mode in the frame of the outflowing
plasma.  Swisdak et al.~\cite{swisdak08a} proposed that the anisotropy
(with typical magnitude $T_x /T_y \sim 2-4$) driving the 
instability arises from cold inflowing plasma mixing with outflow from
the x-line.  Here we apply our analytical results to these
observations. After fitting the inhomogeneity
(Fig.~\ref{weibel_2stream}(d)) seen in the reconnection simulations by
Eqs.~(\ref{inhomo_pic1})-(\ref{inhomo_pic2}) with parameters
$n_h=0.25 n_0, n_b=0.2 n_0, B_{x,h}^2=0.8 B_{x0}^2$ (denoted Profile C)
we can find the necessary temperature anisotropy for Weibel to be
unstable within the current layer.  The (red) dot-dashed curve in
Fig.~\ref{marginal} is the marginal criterion for the Weibel
instability, and it gives us the minimum temperature anisotropy $T_x/T
\sim 3.3$ for a typical reconnection layer of width $4d_e$ (i.e.,
$\delta=2d_e$).  This predicted minimum is within the observed
anisotropy values ($2\sim4$) and thus demonstrates that the
reconnection current layer can support the Weibel mode. The even parity of $B_z$ produced by the Weibel mode (Fig.~\ref{smallrun}(a)) is also seen in the reconnection simulation. Farther
downstream, the nonlinear saturation of the Weibel instability stops
the system from moving to even higher temperature anisotropies and
keeps the system near marginal stability.

The magnitude of $B_z$ saturates at $\sim0.1-0.4B_{x0}$ in
Fig.~\ref{smallrun}(d), which is comparable to the value shown in
Fig.~\ref{weibel_2stream}(b). Therefore by measuring the time,
$t_s$, for $B_z$ to saturate, we can estimate the half-length
of the reconnection nozzle as $L\approx V_{A,e} t_s$ since
$B_z$ develops from the instability while the plasma is convected out
from the x-point at roughly the electron Alfv\'en velocity. As a
result, anisotropy values of $2.5\sim 4$ with saturation times from Fig.~\ref{smallrun}(d) give a predicted
half-nozzle length in the range $25
\sim 80 d_e$, which compares favorably with
the observed reconnection layer half-length of $60 d_e$.  Although
these small runs use higher density plasmas than the reconnection
runs, we expect the saturation behavior to be similar. Furthermore,
the (red) dot-dashed curve in Fig.~\ref{smallrun} (b), which represents the
four-beam model solution with Profile C, indicates that an anisotropy
of $4.0$ produces a growth rate of $\sim 0.2 \Omega_{ce}$, which again leads to a 
nozzle half-length of $\sim50d_e$ (i.e., we expect its
evolution to be similar to the solid curve in Fig.~\ref{smallrun} (d)).

In contrast to our small-box runs where reconnection was suppressed,
we expect the background to be noisier in a simulation that allows
both reconnection and the Weibel instability to develop.  However that
will not strongly affect our results.  Since the out-of-plane field
$B_z$ grows exponentially from the initial noise we expect the
estimated nozzle length to scale logarithmically with the initial
noise level. Hence the predicted
length via this estimate is insensitive to the noise in $B_z$ and we
can expect our small-box runs to still give a reliable comparison to
reconnection simulations.

\subsection{$k_x$ structure}  

\begin{figure}
    \includegraphics[width=8.5cm]{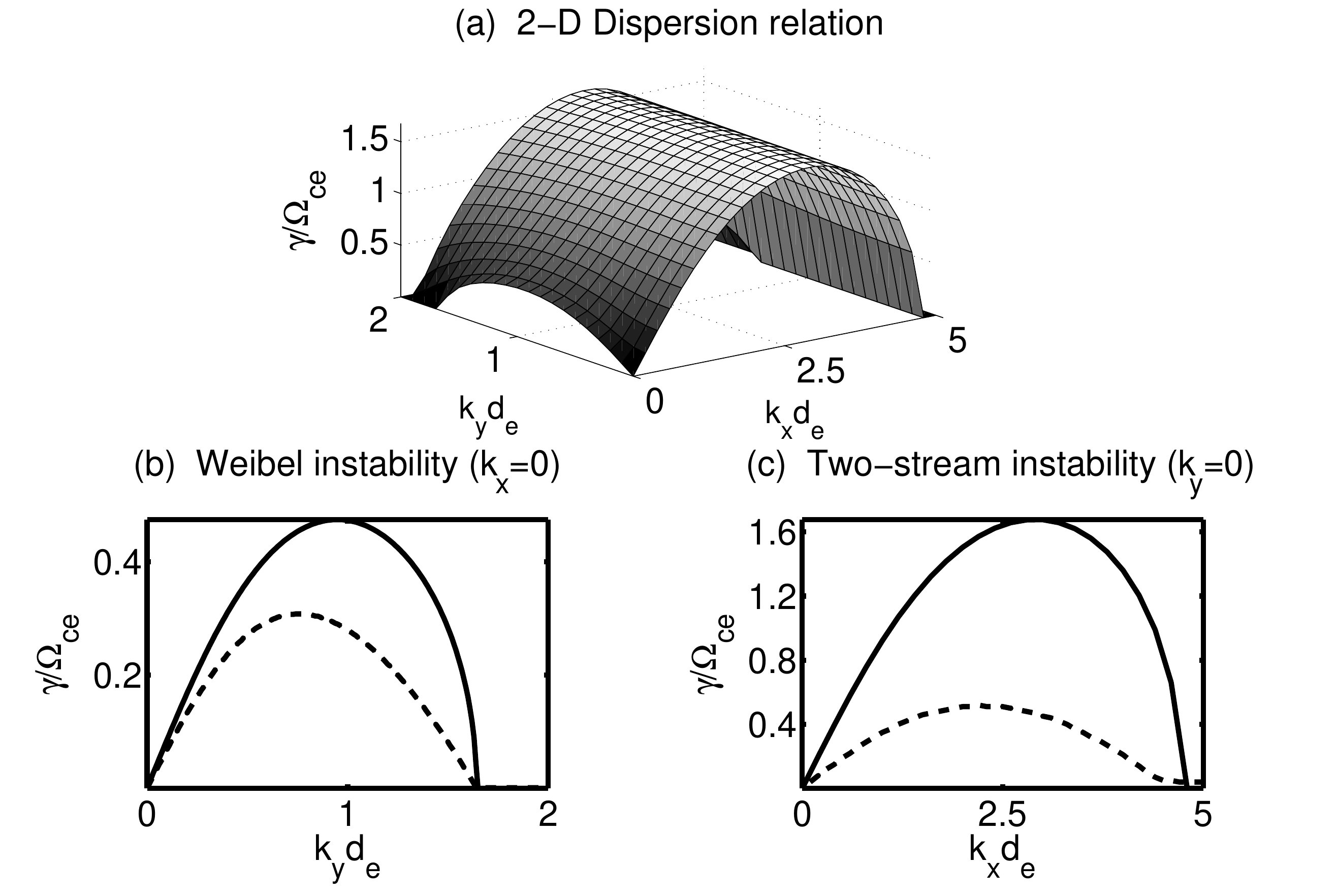} 
\caption{Homogeneous plasma
    dispersion relation with $c=5V_{A,e}, n_{p,e}=0.45n_0, B_x=0,
    T_x=4T_{yy}=4.0 m_eV_{A,e}^2$. In (a) the growth rate as a function of $k_x$ and
    $k_y$.  In (b) the homogeneous
    dispersion relation for the Weibel instability.  The solid curve
    corresponds to the four-beam model and the dashes to kinetic
    theory. In (c) the homogeneous dispersion relation of the
    two-stream instability. The solid curve corresponds to a the
    four-beam model with $T_{xx}=0$. The dashed curve is kinetic
    theory with $T_{xx}=T_{yy}=0.25 m_eV_{A,e}^2$. } \label{twoD}
\end{figure}

We now focus on the origin and magnitude of the finite $k_x$ in the
region downstream of the x-line.  Loosening our assumptions by letting
$\partial/\partial_x \ne 0$ in the linearized four-beam model from Eq.~(\ref{eq_full})-(\ref{eq_full_end}), we can numerically solve for the two-dimensional dispersion relation in a homogeneous plasma.  One new
feature is the introduction of a two-stream instability (i.e., $k_y=0$ mode) that has
previously been noted by Zenitani \& Hesse \cite{zenitani08a} (see Appendix B for the dispersion relation).
As shown in Fig.~\ref{twoD}, the two-stream instability has a
higher growth rate than the Weibel instability. Hence it always grows
in front of the Weibel structures with wavelength $\sim 2\pi/3 \approx
2d_e$, where the factor of 3 comes from the wavenumber for the maximum
growth mode measured in Fig.~\ref{twoD}(c). We do observe a double
peaked distribution of the $x$-direction velocity and the $E_x$
signature of the two-stream instability (Fig.~\ref{weibel_2stream}
(e), (c)) in the appropriate region with a wavelength comparable to the
predicted value. The nonlinear development of the two-stream
instability tends to merge the counter-streaming distributions,
resulting in a single-humped distribution farther downstream with
$T_x>T_y$.  The result is a transition from coexisting Weibel and
two-stream instabilities to a pure Weibel instability, as can be seen
in Fig.~\ref{weibel_2stream}(b).

The two-stream maximum growth rate predicted by the four-beam model is
three times larger than that of the Weibel instability (divide the maximum
growth rate of the solid curve in Fig.~\ref{twoD}(c) by that of
Fig.~\ref{twoD}(b)), which seems inconsistent with their relatively
close development in the reconnection simulations.  However this
disagreement can be reduced by the introduction of kinetic effects and
a finite $T_{xx}$, both of which are not included in our four-beam
model.  In particular, a finite $T_{xx}$ suppresses the growth of the
two-stream instability.  Kinetic theories predict lower growth rates
for both the two-stream and Weibel instabilities, with the ratio of
their fastest growth rates decreasing to about 1.6 (divide the maximum
growth rate of the dashed curve in Fig.~\ref{twoD}(c) by that of
Fig.~\ref{twoD}(b)).  This lower ratio helps to explain the nearly
coincident signatures of both instabilities in pair reconnection. In
general, we also expect that a full kinetic treatment of Harris sheet
inhomogeneous plasmas will produce slightly lower Weibel growth rates
than those of our four-beam model, a feature that has already been
observed in Fig.~\ref{smallrun}(b).

The two-stream instability can not explain the longer $x$-direction
variance of the chess-board-like structure farther downstream. In that
region, $B_x$ decreases from the asymptotic value of the reversed
field ($\approx B_{x0}$) at the nozzle edge to zero at the nozzle symmetry
line while the temperature anisotropy remains large. Although Swisdak
et al. \cite{swisdak08a} use a parity argument to argue against the
possibility that the firehose instability plays a role, it is perhaps
plausible that it could couple to the stronger Weibel mode and provide
the finite $k_x$.  However this possibility is again ruled out by
directly solving for the homogeneous dispersion relation of the
firehose instability via a one fluid double-adiabatic model with
finite Larmor radius corrections \cite{davidson68a}.  The firehose instability in a homogenous plasma with $n_{p,e}=0.45n_0$, $T_x=4.0T$ has the strongest growth
rate $\sim 0.035\Omega_{ce}$ at $B_x \sim 0.44 B_{x0}$ and $k_xd_e \sim 0.07$, which is an
order of magnitude smaller than the growth rates of the Weibel and
two-stream instabilities. Moreover, the predicted wavelength is too
large to explain the observed $k_x$.

\begin{figure}
    \includegraphics[width=8.5cm]{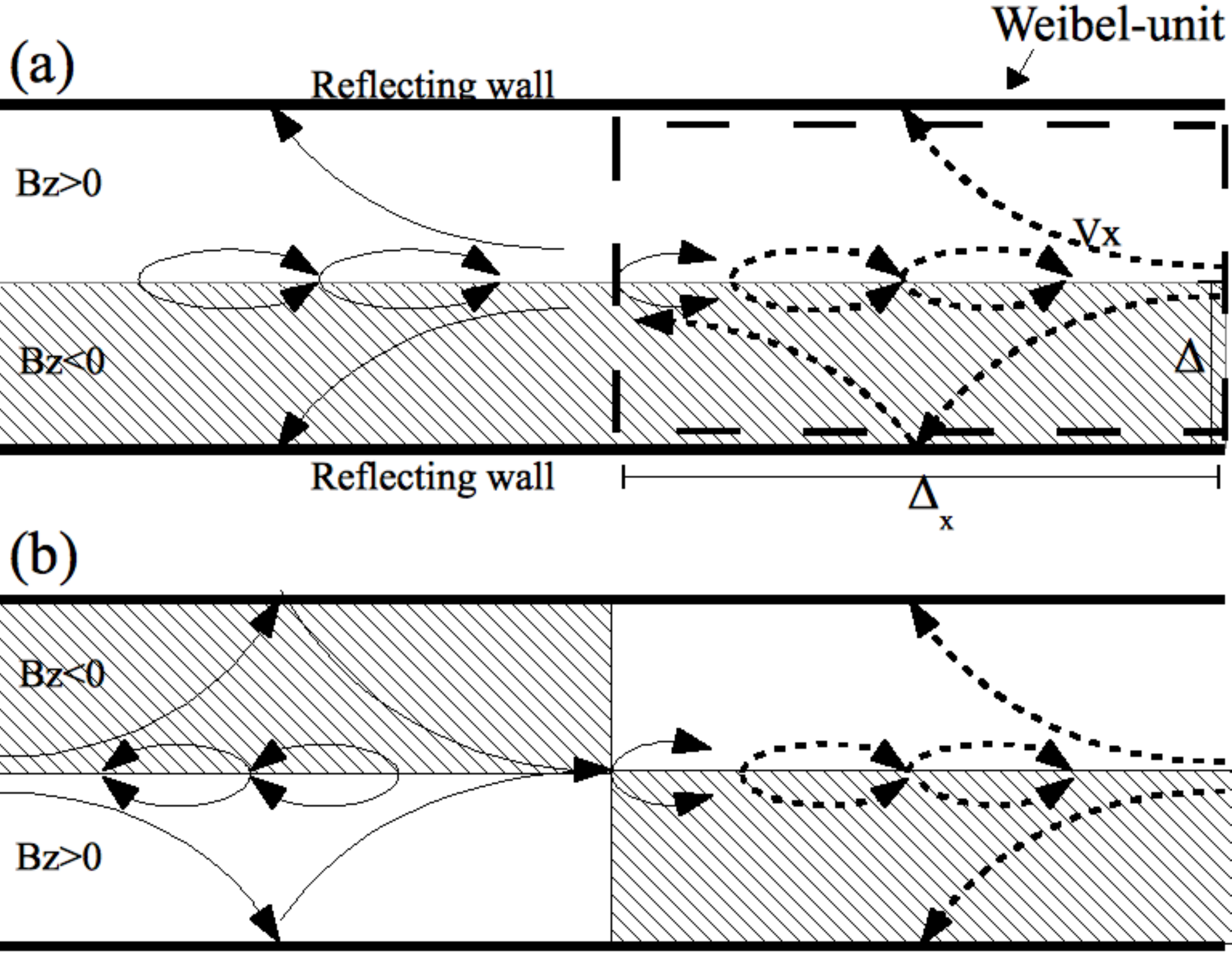} 
\caption{Schematic
    explanation of the interaction between Weibel-units. White regions
    have $B_z>0$, hatched $B_z<0$. A Weibel-unit is represented by particle trajectories similar to the dashed
    curves of Fig.~\ref{weibel_0}. In (a) the converging particles of the right Weibel-unit are replenished by the
    converging particles from the left unit.
    In (b) converging particles of the right Weibel-unit
    can be replenished by diverging particles from the left unit of opposite polarity.}
    \label{weibel_schematic}
\end{figure}

A possible mechanism for the observed $k_x$ is proposed here. First
imagine that the instability is confined in the $y$ direction between
reflecting walls (see Fig.~\ref{weibel_schematic}). The reflecting walls mimic the strong $B_x$ field at the boundaries of the current layer. In this geometry there is an intrinsic scale length $\Delta_x$ associated with the trajectory of a particle away from the current layer, its reflection from the wall, and its motion back towards current layer. For a uniform $B_z$, $\Delta_x=2
\sqrt{\rho_z^2-(\rho_z-\Delta)^2}$, where $\rho_z=V_x/\Omega_{c,z}$ is the
gyro-radius based on $B_z$ and $\Delta$ is the half distance between the walls. 
The particle trajectory is sketched in the dashed box of Fig.~\ref{weibel_schematic}(a). We suggest that this intrinsic scale length controls the x-dependence of $B_z$ seen in the simulations.
We define the scale $\Delta_x$ shown as a dashed box in Fig.~\ref{weibel_schematic}(a) as a Weibel-unit
and consider the interaction between two Weibel-units.
When all of the converging streaming particles (dashed curves with rightward arrows) leave the center
of the righthand Weibel-unit in Fig.~\ref{weibel_schematic}(a), this unit will need a replenishment of converging streaming particles (solid curves with rightward arrows) from its leftward
neighbor to maintain its central current. 
Hence this self-consistent arrangement can arise without x-variation.

However, suppose that due to the initial random noise the lefthand
Weibel-unit acquires the opposite polarity magnetic field, as seen in
Fig.~\ref{weibel_schematic}(b). Then the intrinsic $\Delta_x$ will
arise such that the source for replenishing the converging streaming
particles of the righthand Weibel-unit is the diverging streaming
particles (solid curves with rightward arrows) of a leftward neighbor after they bounce
(perhaps multiple times) against the walls. The lefthand Weibel-unit
can be reinforced from its rightward neighbor in the same
manner. In this configuration we can estimate the distance between two
neighboring Weibel-units to be $N\Delta_x$, where $N$ is the number of times a particle reflects from the
walls.

\begin{figure}
    \includegraphics[width=8.5cm]{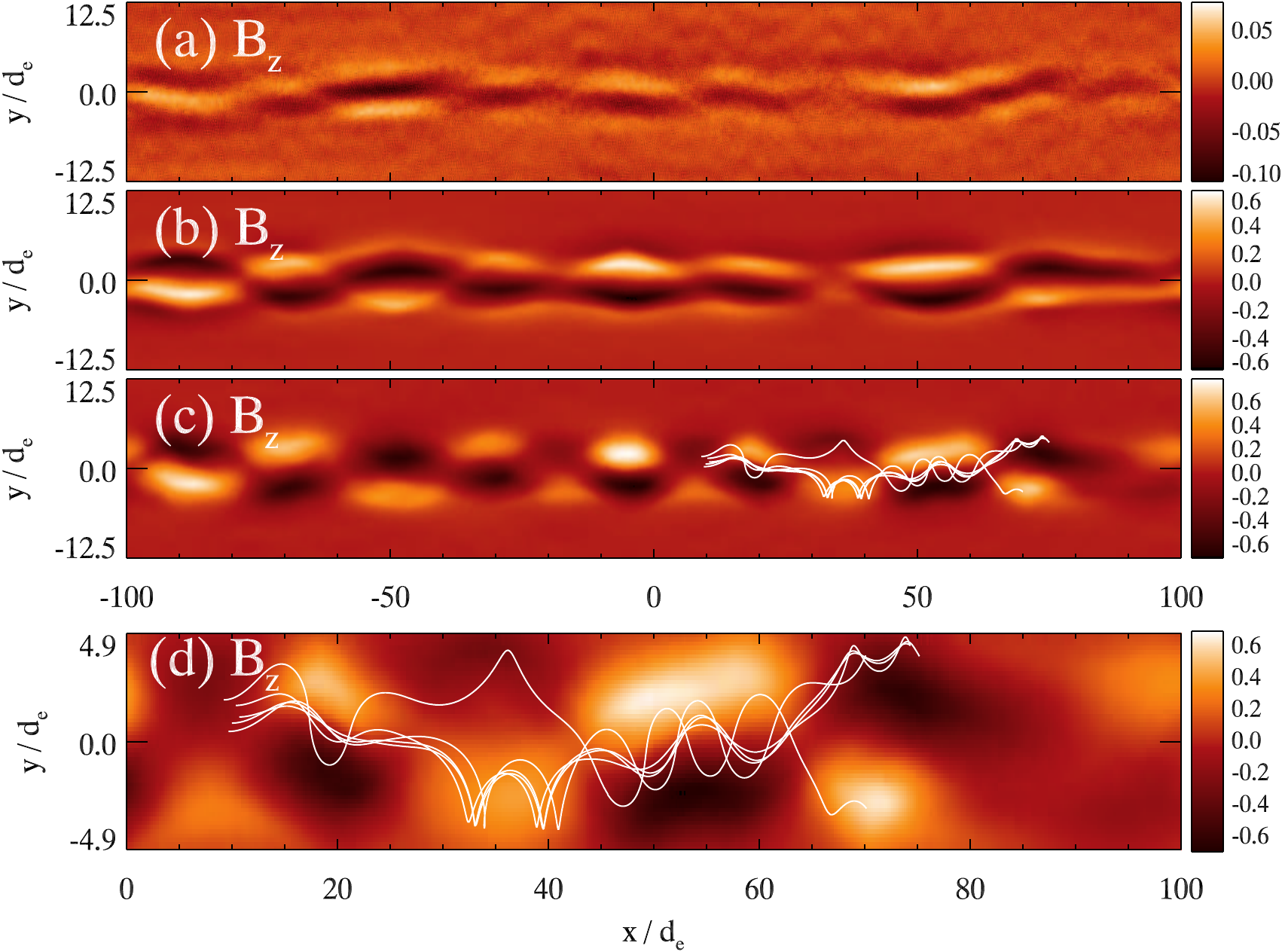} 
\caption{(Color online) Evolution of the
    Weibel instability with initial anisotropy $4.0$ inside a magnetic
    field trough. In (a)--(c), $B_z$ at
    $t\Omega_{ce}=12.5,25,37.5$. White curves mark the trajectories of
    five test positrons with initial velocity $\sqrt{T_x/m}=1.0V_{A,e}$ and initial position ($10d_e,0d_e$). In
    (d) enlargement of the particle trajectories at time
    $37.5/\Omega_{ce}$.}
     \label{kx_evolve} 
    \end{figure}

This qualitative explanation is supported by PIC
simulations, as can be seen in Fig.~\ref{kx_evolve}. We confine an
plasma with an anisotropic temperature within a magnetic trough with
thickness $5d_e$.  Specifically, $B_x=0$, $T_x=4.0T$, and $n=1.2n_0$ between
$y=-2.5 \sim 2.5d_e$ and $B_x=1.0B_{x0}$, $T_x=T$, and $n=0.2n_0$ outside this
region. The simulation is performed in a domain of size $200d_e \times
25d_e$ with periodic boundary conditions in both the $x$ and $y$
directions. Here the $B_x$ trough serves as the reflecting walls and
the trough thickness of $5d_e$ is comparable to the reconnection
nozzle thickness. Two distant Weibel-units are not able to communicate
with each other, and therefore it is not surprising to see that the
magnetic fields at $x=-50d_e$ and $50d_e$ in Fig.~\ref{kx_evolve}(a)
have opposite signs.  As time evolves, the Weibel-generated field gets
stronger while the long Weibel-unit ($x=-30d_e \sim 30d_e$ at
Fig.~\ref{kx_evolve}(b)) breaks up into smaller Weibel units of
opposite polarity at (c). Five test positrons with initial velocity
$\sqrt{T_x/m}=1.0 V_{A,e}$ (i.e., $T_x/T=4$) are randomly placed near $(10d_e,
0 d_e)$. Their trajectories are shown as white curves in
Fig.~\ref{kx_evolve}(c) and are blown up in Fig.~\ref{kx_evolve}(d).
They are qualitatively similar to those described in
Fig.~\ref{weibel_schematic}(b). The converging motions of those
trajectories at the trough center between $x=10\sim30d_e$ and
$x=45\sim65d_e$ of Fig.~\ref{kx_evolve}(d) contribute to the current
for the out-of-plane magnetic field. These particles reflect from
$B_x$ in between $x=30\sim45d_e$, where they deposit their momenta. At late
time, $t\Omega_{ce}=37.5$, the entire channel relaxes to a chain of
Weibel units of alternating polarity, whose length scale is approximately determined by $N\Delta_x$ with $N\gtrsim 1$.

In order to apply this idea to the downstream regions in reconnection
simulations, we approximate the characteristic streaming velocity as
$1.0 V_{A,e}$ and the averaged out-of-plane magnetic field as $0.2 B_{x0}$.  This
implies a gyro-radius of $5d_e$. The layer thickness $\Delta$ is
$2d_e$ along the reconnection nozzle, and thus $\Delta_x$ has a scale of approximately $2\sqrt{5^2-(5-2)^2}=8d_e$,
about half of the size of the observed structure in
Fig.~\ref{weibel_2stream}(c). This mechanism roughly explains the scale
size of the observed variation. Furthermore, the gyromotion of the
reconnection outflow helps explain how the reconnection nozzle
broadens downstream. The out-of-plane magnetic field bends the outflow
momentum from the $x$ to the $y$-direction. As a result this flow
pushes the background Harris magnetic field away from the
symmetry line, broadening the current layer.

\section{Summary and Discussion}\label{secsumm}

We have developed a four-beam model to study the effects on the Weibel
instability of a spatial inhomogeneity arising from a current layer. We have shown that the Weibel instability is able to grow
within narrow Harris sheets, and its growth rate (saturation time),
saturation magnitude, and mode structure fit those values observed in
pair reconnection.

This further suggests that the Weibel instability might control the current layer
dynamics in pair reconnection, where the Hall term is absent. Other
candidate instabilities, such as the two-stream and firehose
instabilities have rather minor effects, particularly since it is not
clear if the firehose instability even appears in these systems.  The
high-velocity outflow scatters into the transverse direction due to the Lorentz force
arising from the Weibel-generated out-of-plane magnetic field. We
argue that the associated increased pressure $P_{yy}$
is responsible for opening the pair reconnection nozzle and shortening the current layer. As a consequence, the shorter current layer generates a higher reconnection rate.

Even though a similar temperature anisotropy also could arise in
electron-proton reconnection, signatures of the Weibel instability are
not seen there. The reason is that the even shorter current layer ($\sim 10
d_e$), controlled by whistler waves \cite{mandt94a}, leaves insufficient space for the
unstable Weibel mode to grow.  It is an open question as to how the
Weibel instability controlled reconnection transforms into whistler
mediated reconnection as the electron to ion mass ratio changes.

The development of the Weibel instability in the initial state of
relativistic ($k_BT\sim mc^2$) pair reconnection is described in
Zenitani \& Hesse \cite{zenitani08a}.  There the Weibel instability is shown to grow in
front of a tangential discontinuity formed by the mixing of outflowing
and ambient plasma, but it is not clear if this turbulence plays a
role in the steady-state development of the outflow exhaust and the associated current layer.  However, by
artificially suppressing the Weibel-generated out-of-plane magnetic
field, they do demonstrate the ability of the Weibel mode to broaden
the current layer, similar to the behavior of the present non-relativistic case.
The growth rate and wave vector of the relativistic Weibel instability
are smaller by a factor of $\gamma_L^{1/2}$, where $\gamma_L$ is the
Lorentz factor \cite{yoon87a}. Consequently, we expect the
instability to grow more slowly during relativistic pair reconnection,
not only because of the intrinsically lower growth rate but also
because of the relatively larger suppressing effect of the Harris
reversed field on the enlarged mode structure (if we assume a similar
nozzle thickness in both the relativistic and non-relativistic
regimes). Overall, it remains an open question as to whether the Weibel
instability plays an important role in controlling relativistic electron-positron reconnection. 

\acknowledgments

Y.\ -H.\ L.\ acknowledges helpful discussions with Dr.\ P.\ N.\ Guzdar. This work was supported in part by NSF grant ATM0613782. Computations were carried out at the National Energy Research
Scientific Computing Center.

\section{Appendix A: The derivation of the governing equations (\ref{eq1}) and (\ref{eq2})}

To derive Eqs.~(\ref{eq1}) and (\ref{eq2}), we define the variables,
\begin{equation}
\tilde{\zeta} \equiv \tilde{V}_{x,p+}+\tilde{V}_{x,p-}-\tilde{V}_{x,e+}-\tilde{V}_{x,e-},
\end{equation}
\begin{equation}
\tilde{\chi} \equiv \tilde{V}_{y,p+}-\tilde{V}_{y,p-}-\tilde{V}_{y,e+}+\tilde{V}_{y,e-},
\end{equation}
\begin{equation}
\tilde{\vartheta} \equiv \tilde{V}_{z,p+}-\tilde{V}_{z,p-}+\tilde{V}_{z,e+}-\tilde{V}_{z,e-},
\end{equation}
\begin{equation}
\tilde{\eta} \equiv \tilde{n}_{p+}-\tilde{n}_{p-}-\tilde{n}_{e+}+\tilde{n}_{e-},
\end{equation}

then combine Eqs.~(\ref{linear1})-(\ref{linear6}) to yield
\begin{equation}
-\gamma \tilde{\eta}=n'\tilde{\chi}+n\tilde{\chi}',
\label{linear11}
\end{equation}
\begin{equation}
\gamma m \tilde{\zeta} = 4 e \tilde{E_x}-mV_x' \tilde{\chi},
\label{linear22}
\end{equation}
\begin{equation}
\gamma m \tilde{\chi} =\frac{e}{c}B_x\tilde{\vartheta} -\frac{4 e}{c} V_x\tilde{B_z}+T\frac{n'}{n^2}\tilde{\eta}-\frac{T}{n}\tilde{\eta}',
\label{linear33}
\end{equation}
\begin{equation}
\gamma m \tilde{\vartheta} = -\frac{e}{c} B_x \tilde{\chi},
\label{linear44}
\end{equation}
\begin{equation}
\tilde{B_z}' = \frac{4\pi e}{c} V_x \tilde{\eta}+\frac{4 \pi e}{c} n \tilde{\zeta},
\label{linear55}
\end{equation}
\begin{equation}
\gamma \tilde{B_z} = c \tilde{E_x}'.
\label{linear66}
\end{equation}
Eqs.~(\ref{linear11})-(\ref{linear66}) represent a set of $6$ equations for the $6$ variables, $\{ \tilde{\eta}, \tilde{\zeta}, \tilde{\chi}, \tilde{\vartheta}, \tilde{B_z}, \tilde{E_x} \}$. Note that $\{ n, V_x, B_x, T\}$ are unperturbed quantities specifying the initial conditions. Now use Eqs.~(\ref{linear11}), (\ref{linear22}), and (\ref{linear66}) to rewrite Eq.~(\ref{linear33}) in terms of $\tilde{\chi}$ and $\tilde{E_x}$,

\begin{equation}
\gamma^2 \tilde{\chi} = -\Omega^2 \tilde{\chi}-\frac{4e}{m}V_x \tilde{E_x}'+C_s^2\left( \frac{(n\tilde{\chi})'}{n}\right)'.
\label{eq11}
\end{equation}

Use Eqs.~(\ref{linear11}), (\ref{linear44}), and (\ref{linear66}) to rewrite Eq.~(\ref{linear55}) in terms of $\tilde{\chi}$ and $\tilde{E_x}$,

\begin{equation}
\tilde{E_x}'' = -\frac{4 \pi e}{c^2} V_x (n \tilde{\chi})'+\frac{2}{d^2} \tilde{E_x}-\frac{4 \pi e}{c^2} V_x' n \tilde{\chi}.
\label{eq22}
\end{equation} \newline 

After some minor algebraic manipulations, Eqs.~({\ref{eq11}}) and ({\ref{eq22}}) can be written in the form shown in Eq.~({\ref{eq1}}) and ({\ref{eq2}}).

\section{Appendix B: The dispersion relations of the two-stream instability and the Weibel instability}
If we consider the $k_y=0, k_x \neq 0$ limit of our four-beam model, we arrive at the usual two-stream instability dispersion relation,

 \begin{equation}
1=\frac{\omega_{p}^2}{(k_xV_x-\omega)^2}+\frac{\omega_{p}^2}{(k_xV_x+\omega)^2},
 \end{equation}
where $\omega= \omega_r+i\gamma$.

The kinetic version of this dispersion relation with finite $T_{xx}$ is
 \begin{equation}
 \ k_x^2 C_s^2+\omega_p^2[2+\xi_1 Z(\xi_1)+\xi_2 Z(\xi_2)]=0,
 \end{equation}
where $Z(\xi) \equiv (1/\sqrt{\pi}) \int_{-\infty}^{\infty} \mbox{exp}(-x^2)/(x-\xi) dx$,
$\xi_1 \equiv ( \omega/|k_x|-V_x)/\sqrt{2T_{xx}/m}$ and
$\xi_2 \equiv (\omega/|k_x|+V_x)/\sqrt{2T_{xx}/m}$.

For reference, the dispersion relation for the Weibel instability ($k_x=0, k_y \neq 0$) in kinetic theory is
\begin{equation}
k_y^2c^2-\omega^2+2\omega_p^2\left(1-\frac{T_x}{T_y}\right)=2\omega_p^2\frac{T_x}{T_y}\xi
Z(\xi),
\end{equation}
where 
$\xi \equiv \omega/(|k_y| \sqrt{2T_y/m})$.
Note that this reduces to Eq.~(\ref{Krall}) in the strong anisotropy limit.

If both $k_x$ and $k_y \neq 0$, both instabilities are present. We treat this limit numerically because of the complexity.


\newpage
\end{document}